\documentclass[a4paper,11pt]{article}
\usepackage{pos}

\usepackage{graphicx}
\usepackage{epsfig}
\usepackage{amsmath, amsthm}
\usepackage{amsfonts}
\usepackage{color}
\usepackage{xspace}
\usepackage{paralist}

\title{Precise predictions for electroweak $t\bar t$ production at the LHC in
models with flavour non-diagonal $Z'$ boson couplings and $W'$ bosons}
\ShortTitle{Electroweak $t\bar t$ production in models with flavour non-diagonal $Z'$ couplings and $W'$s}

\author*[a,c,1,2]{M.M.~Altakach}
\author[b,3]{T.~Je\v{z}o}
\author[c,2,4]{M.~Klasen}
\author[d,5]{J.-N.~Lang}
\author[a,1]{I.~Schienbein}

\note{Work supported by the IN2P3project ``Th\'eorie – BSMG''.} 
\note{Work supported by the BMBF under contract 05H18PMCC1 and by the DAAD.} 
\note{Work supported by the DFG under grant 396021762 - TRR257.} 
\note{Work supported by the BMBF under contract 05H18PMCC1.} 
\note{Work supported by the SNSF under contract BSCGI0-157722.} 
\note{Preprint numbers: KA-TP-19-2020, MS-TP-20-49, P3H-20-067, ZU-TH-53/20}

\affiliation[a]{Laboratoire de Physique Subatomique et de Cosmologie,
 Universit\'e Grenoble-Alpes, CNRS/IN2P3,
 53 Avenue des Martyrs, 38026 Grenoble, France}

\affiliation[b]{Institut f\"ur Theoretische Physik, Karlsruhe Institut f\"ur Technologie, 76128 Karlsruhe, Germany}

 \affiliation[c]{Institut f\"ur Theoretische Physik, Westf\"alische
 Wilhelms-Universit\"at M\"unster, Wilhelm-Klemm-Stra\ss{}e 9, 48149
 M\"unster, Germany}

\affiliation[d]{Physik-Institut, Universit\"at Z\"urich, 8057 Z\"urich, Switzerland}

\emailAdd{altakach@lpsc.in2p3.fr}

\abstract{
We report on our re-calculation of electroweak top-quark pair production in
Standard Model extensions with extra heavy neutral and charged spin-1 particles
at the LHC with substantial improvements.
In particular, we allow for flavour--non-diagonal $Z'$ couplings and take into
account non-resonant production in the Standard Model and beyond.
As in our previous work we include NLO QCD corrections and match to parton
showers with the POWHEG method fully taking into account the Standard Model and
new physics interference effects.
We consider the Sequential Standard Model, the Topcolour model as well as the Third
Family Hypercharge Model featuring non-flavour diagonal $Z'$ couplings which
has been proposed to explain the anomalies in $B$ decays.
Numerical results for $t\bar{t}$ cross sections at hadron colliders with
$\sqrt{s}$ of up to 100 TeV are presented. 
We also investigate the numerical impact of the new non-resonant contributions.
}

\FullConference{%
  40th International Conference on High Energy physics - ICHEP2020\\
  July 28 - August 6, 2020\\
  Prague, Czech Republic (virtual meeting)
}

\begin{document} 
\maketitle

\section{Introduction\label{sec:intro}}
The Standard Model (SM) of particle physics, based on a $SU(3)_C \times SU(2)_L
\times U(1)_Y $ gauge symmetry, is an extremely successful theory accounting
for a wide range of high energy experiments at both the intensity and energy
frontiers.
Nevertheless, it is widely believed to be incomplete for multiple reasons: it
doesn't provide a candidate for a dark matter particle; the CP violation in the
SM is not sufficient to explain the observed matter--anti-matter asymmetry; and
massive neutrinos are, a priori, not accounted for.
Furthermore, its scalar sector is plagued by naturalness problems related to
the stability of the Higgs mass under quantum correction and to the suppression
of CP violation in the strong interaction.
Hence, the general expectation has been for a long time that new
physics beyond the SM should be present close to the electroweak (EW) scale. 

Despite the fact that no signals of new physics were found in the first two
runs of the Large Hadron Collider (LHC) at CERN, there are still high hopes
that new particles will show up in future high-luminosity runs.
Such signals will likely appear as small deviations from the SM predictions
which makes precise predictions for both the SM background and the new physics
signals increasingly important.

We focus on new heavy electrically charged or neutral spin-1 resonances
with EW-like couplings, denoted ${W'}^\pm$ and $Z'$,
respectively. 
They are predicted by several well-motivated extensions of the SM and are
extensively looked for at the LHC.
In this context it is noteworthy that $Z'$ models with a non-universal flavour
structure \cite{Allanach:2018lvl, Allanach:2019iiy, Crivellin:2015mga, Crivellin:2015lwa}, where the $Z'$ couples
differently to the fermions of the three SM families, are viable candidates to
explain the current B-flavour anomalies \cite{Aaij:2017vbb, Aaij:2015oid,
Aaij:2013qta, Aaij:2017vad, Aaboud:2018mst, CMS:2014xfa, Chatrchyan:2013bka,
Bobeth:2017vxj, Khachatryan:2015isa,Alguero:2019ptt}.

In many cases, the strongest constraints on the parameter space of models with
$Z'$ and $W'$ resonances come from searches with dilepton final states.
In this case, precise predictions at next-to-leading order (NLO) accuracy
including a resummation of soft gluon terms at next-to-leading logarithmic
accuracy can be obtained with the {\tt Resummino} code \cite{Fuks:2013vua,
Jezo:2014wra, CidVidal:2018eel}.
However, top quark observables are also very interesting since the 3rd
generation plays a prominent role in the SM due to the Yukawa coupling of the
top quark, which is the only Yukawa coupling in the SM which is of order one.
Therefore, it is quite conceivable that new physics couples predominantly to
the top quark.

In 2015, some of us performed a calculation of NLO QCD corrections to the EW
$t\bar{t}$ production in the presence of a $Z'$ resonance matched to parton
showers (PS) with the POWHEG method \cite{Bonciani:2015hgv}.
In this proceedings we report on a complete re-calculation including a number
of improvements, among which the most noteworthy are: (a) the generalisation of
$Z'$ couplings also allowing for flavour--non-diagonal fermion vertices; (b)
addition of non-resonant contribution such as the $t$-channel $W$ and $W'$
exchange. 
In Sec.~\ref{sec:calculation} we provide a description of the calculation with
a focus on those aspects which differ from our previous calculation in
\cite{Bonciani:2015hgv}. 
Then in Sec.~\ref{sec:models} we briefly summarise the models considered here. 
The tool setup is described and the numerical results are presented in
Sec.~\ref{sec:numerics}.
Finally, in Sec.\ \ref{sec:conclusions} we present a summary and our
conclusions.

\section{Description of the calculation\label{sec:calculation}}
The cross section for the hadroproduction of a $t \bar t$ pair, $AB \to t \bar
t X$, is given by the usual convolution of the parton distribution functions
(PDFs) inside the two incoming hadrons $A$ and $B,$ with the short distance
cross sections $\hat \sigma_{ab}$ summed over partonic channels, $ab \to t \bar t
X$.
Here we work in a 5-Flavour Number Scheme including all relevant contributions
with $u, d, s, c, b$ (anti-)quarks, gluons and unless explicitly stated also
photons in the initial state.

Up to NLO the hard scattering cross sections have the following perturbative
expansion in the  strong ($\alpha_s$)  and electroweak
($\alpha$) coupling constants:
\begin{equation}
\hat \sigma=
\hat \sigma_{2;0}(\alpha_{S}^2)
 + \hat \sigma_{3;0}(\alpha_S^3)
 + \hat \sigma_{2;1}(\alpha_S^2\alpha)
 + \boldsymbol{\hat \sigma_{0;2}(\alpha^2)}
 + \boldsymbol{\hat \sigma_{1;2}(\alpha_{S}\alpha^2)}
 + \boldsymbol{\hat \sigma_{1;1}(\alpha_{S}\alpha)}
 + \hat \sigma_{0;3}(\alpha^3) \ ,
 \label{eq:1.2}
\end{equation}
where the numerical indices $(i;j)$ in $\hat \sigma_{i;j}$ represent the powers
in $\alpha_s$ and $\alpha$, respectively, the parton flavour indices and the
dependence on the scales have been suppressed, and the terms considered in our
calculation have been highlighted in bold.

In our calculation we focus on the tree-level EW top-quark pair production,
$\boldsymbol{\hat \sigma_{0;2}}$, and its NLO QCD corrections,
$\boldsymbol{\hat \sigma_{1;2}}$.
$\boldsymbol{\hat \sigma_{0;2}}$ receives contributions from the $s$-channel
amplitudes $q \bar{q} \to Z',Z,\gamma \to t \bar t$, including the $Z'$ signal
and its interference with the photon and SM $Z$ boson.
Due to the resonance of the $Z'$ boson, we expect these terms to be the most
relevant for new physics searches.
In addition, we include new contributions from diagrams with non-resonant
exchanges of $W$, $W'$ and $Z'$ bosons that were not considered in
Ref.~\cite{Bonciani:2015hgv}.
Note that out of these, the first two take into account CKM mixing and the last
one is only allowed in models with flavour non-diagonal couplings.
A particular advantage of the $\boldsymbol{\hat \sigma_{0;2}}$ contribution is
that the calculation of $\boldsymbol{\hat \sigma_{1;2}}$ can then be carried
out in a model-independent way.
We also consider the term $\boldsymbol{\hat \sigma_{1;1}}$ that includes both
the contribution from the photon induced subprocess and the previously not
considered interference of the $s$-channel QCD and the $t$-channel EW top-pair
production\footnote{That is the interference of $q\bar{q} \to g \to t \bar t$
and $q \bar{q} \to W^*,W'^*,Z'^* \to t \bar t$, where ${}^*$ indicates
$t$-channel exchange.}.
The photon induced subprocess, $\gamma g \to t \bar t$, is included for a
consistent treatment of the mass singularities in the process $gq  \to t \bar t
q$ when the $t$-channel photon is collinear to the quark and is numerically
important.
However, we neglect photon-initiated contributions to $\boldsymbol{\hat
\sigma_{0;2}}$ and $\boldsymbol{\hat \sigma_{1;2}}$.

We do not consider any of the remaining terms in Eq.~(\ref{eq:1.2}).
The terms $\hat \sigma_{2;0}$ and $\hat \sigma_{3;0}$ do not get affected by
the presence of $Z'$ or $W'$ bosons and are readily available in many NLO+PS
event generators \cite{Frixione:2007nw, Campbell:2014kua, Jezo:2016ujg,
Gleisberg:2008ta, Alwall:2014hca, Bellm:2015jjp}.
The EW corrections in terms $\hat \sigma_{2;1}$ and $\hat \sigma_{0;3}$ would 
receive corrections from new $Z'$ and $W'$ running in the loops. 
However, they are parametrically suppressed relative to their underlying Born
processes $\hat \sigma_{2;0}$ and $\hat \sigma_{0;2}$\footnote{Despite the
parametric suppression they have been shown to be important in the SM in
regions with large top transverse momentum \cite{Pagani:2016caq,
Gutschow:2018tuk}.} and, unfortunately, do not lend themselves to model
independent calculations.

Born, virtual and real amplitudes are calculated using the public library {\tt
Recola2}, an extension of {\tt Recola} \cite{Actis:2016mpe} for the computation
of tree and one-loop amplitudes in the SM and beyond.\footnote{The model file
used here can be found on the official website
\url{https://recola.hepforge.org/} under model files. }
We implement calls to {\tt Recola} subroutines in the {\tt POWHEG\,BOX} event
generator \cite{Alioli:2010xd} that allows matching NLO calculations to PS
(jointly referred to as NLO+PS) using the POWHEG method \cite{Nason:2004rx,
Frixione:2007vw}.

\section{Models\label{sec:models}}
We consider three extensions of the SM: the Sequential Standard Model (SSM)
\cite{Altarelli:1989ff}, the Top Colour model (TC) and the Third Family
Hypercharge Model (TFHM) \cite{Allanach:2018lvl}.
The SSM is a toy model which introduces copies of the $W$ and $Z$
bosons that only differ in mass.
It has no free parameters except for the $Z'$ and $W'$ masses.
Due to its simplicity and convenience it is a widely used benchmark model in
which LHC data are analyzed.
In the TC model, which can generate a large top-quark mass through the
formation of a top-quark condensate, the $Z'$ boson is leptophobic and couples
only to the quarks of the first and the third generation.
It has three parameters: the ratio of the two $U(1)$ coupling constants,
$\cot{\theta_H}$, which should be large to enhance the condensation of top
quarks, but not bottom quarks, as well as the relative strengths $f_1$ and
$f_2$ of the couplings of right-handed up- and down-type quarks with respect to
those of the left-handed quarks.
THFM extends the SM by an anomaly-free, spontaneously-broken U(1)$_F$ gauge
symmetry.
Apart from the new gauge boson and a SM singlet, complex scalar field, needed
for the spontaneous symmetry breaking of the U(1)$_F$ symmetry, no new
particles are introduced. 
The model has flavour-dependent couplings and provides an explanation of the
heaviness of the third generation of SM particles and the smallness of the
quark mixing. 
Recently, the TFHM has been slightly modified to make it more natural in the
charged lepton sector \cite{Allanach:2019iiy}, however, in the following we
will use the original TFHMeg from Ref.~\cite{Allanach:2018lvl,
Allanach:2019mfl}. 
It has three free parameters: the extra U(1) coupling, $g_F$, the angle
controlling the mixing of the second and third family quarks, $\theta_{sb}$ and
the $Z'$ boson mass.

The most stringent exclusion limits for the SSM are derived from searches with
dilepton and charged lepton plus missing transverse momentum final states.
They exclude $Z'$ and $W'$ with masses respectively below $5.2$ and $6.0$ TeV 
\cite{Aad:2019fac, CMS:2019tbu, Aad:2019wvl, Sirunyan:2018mpc, Sirunyan:2018lbg}.
The TC model is excluded with $Z'$ masses below $3.8-6.65$ TeV in $t\bar{t}$ channel 
\cite{Sirunyan:2018ryr, Aad:2020kop}.
The parameter space of the THFMeg model has been investigated in
\cite{Davighi:2019jwf} and for the choice $\theta_{sb} = 0.095$, $g_F/m_{Z'} =
0.265$ roughly corresponds to the range $m_{Z'} \in (2,5)$ TeV.
The intention of this work is to demonstrate the variety of SM extensions that
our calculation can be applied to. 
Therefore, these exclusion limits are disregarded in the comparisons that follow.

\section{Numerical results\label{sec:numerics}}
We present NLO+PS predictions for EW top-pair production for a $pp$ collider
with a range of collider energies $\sqrt{s} \in \{14, 27, 50, 100\}$ TeV.
We consider the SSM, TC and THFM models and a range of $Z'$ masses $m_{Z'} \in
[2,8]$ TeV.
In the SSM we set the mass of $W'$ equal to the mass of $Z'$. 
The widths of $Z'$ and $W'$ bosons must then be $\Gamma_{Z'}/m_{Z'} =
3\%$, $\Gamma_{W'}/m_{W'} = 3.3\%$.
The parameters of the TC model are chosen as follows: we set $f_1 = 1$ and $f_2 = 0$
and calculate $\cot \theta_H$ such that $\Gamma_{Z'}/m_{Z'} = 3.1-3.2 \%$.
In THFM we set $\theta_{sb} = 0.095$, $g_F/m_Z' = 0.265$ which implies
$\Gamma_{Z'}/m_{Z'} = \{0.012, 0.028, 0.050, 0.078, 0.112, 0.152\}$ for
$m_Z' = \{2, 3, 4, 5, 6, 7\}$ TeV.

We employ a top quark pole mass $m_t = 172.5$ GeV.
Furthermore, the masses and widths of the weak gauge bosons are given by $m_Z =
91.1876$ GeV, $\Gamma_Z = 2.4952$ GeV, $m_W = 80.385$ GeV, $\Gamma_W = 2.085$
GeV \cite{Patrignani:2016xqp}.
The weak mixing angle is fixed by  $\sin^2 \theta_W = 1-m_W^2/m_Z^2 = 0.222897$
and the fine-structure constant is set to  $\alpha(2 m_t) = 1/126.89$.  We
neglect the running of this coupling to higher scales.

For the proton PDFs, we use the NLO luxQED set of NNPDF3.1
\cite{Bertone:2017bme, Manohar:2016nzj, Manohar:2017eqh} as implemented in the
LHAPDF library (ID = 324900) \cite{Buckley:2014ana, Andersen:2014efa} both at
NLO+PS and at LO+PS.
This set provides, in addition to the gluon and quark PDFs, a precise
determination of the photon PDF inside the proton which we need for our cross
section predictions.
The running strong coupling $\alpha_s(\mu_R)$ is evaluated at NLO in the
$\overline{\text{MS}}$ scheme and is provided together with the PDF
set\footnote{Its value is fixed by the condition $\alpha_s(m_Z) = 0.118$.}.

For our numerical predictions we choose equal values for the factorisation and
renormalisation scales, $\mu_F$ and $\mu_R$ respectively, which we identify
with the partonic centre-of-mass energy: $\mu_F = \mu_R = \sqrt{\hat s}$.

We generate events in the Les Houches Events format \cite{Alwall:2006yp} using
{\tt POWHEG\,BOX} with stable on-shell top quarks requiring the underlying Born
kinematics to satisfy a cut on the $t \bar t$ invariant mass, $m_{t \bar t} \ge
0.75 m_{Z'}$, in order to enhance the signal over background ratio.
We then decay both top quarks leptonically and shower the events using {\tt
PYTHIA\,8.2} \cite{Sjostrand:2007gs}.
The branching ratio of the leptonic top decay of $10.5\%$ \cite{Zyla:2020zbs}
squared is applied.
We use {\tt PowhegHooks} to veto shower emissions harder than the POWHEG 
emission and disable QED shower emissions.

We perform further event selection and histogram on the fly
using {\tt Rivet} \cite{Buckley:2010ar, Bierlich:2019rhm}.
Events are required to have two or more charged leptons, two or more neutrinos and 
two or more anti-$k_T$ \cite{Cacciari:2008gp} $R=0.5$ jets, each containing at
least one $b$-parton.
All these objects have to fulfil the acceptance cuts $p_T > 25$ GeV and
$|y|<2.5$.
Furthermore we combine charged leptons and neutrinos into $W$ bosons based on
their MC truth PDG id and require each event to feature at least one such $W^+$
and one such $W^-$ boson.

\begin{figure}
 \centering
   \includegraphics[width=0.45\textwidth]{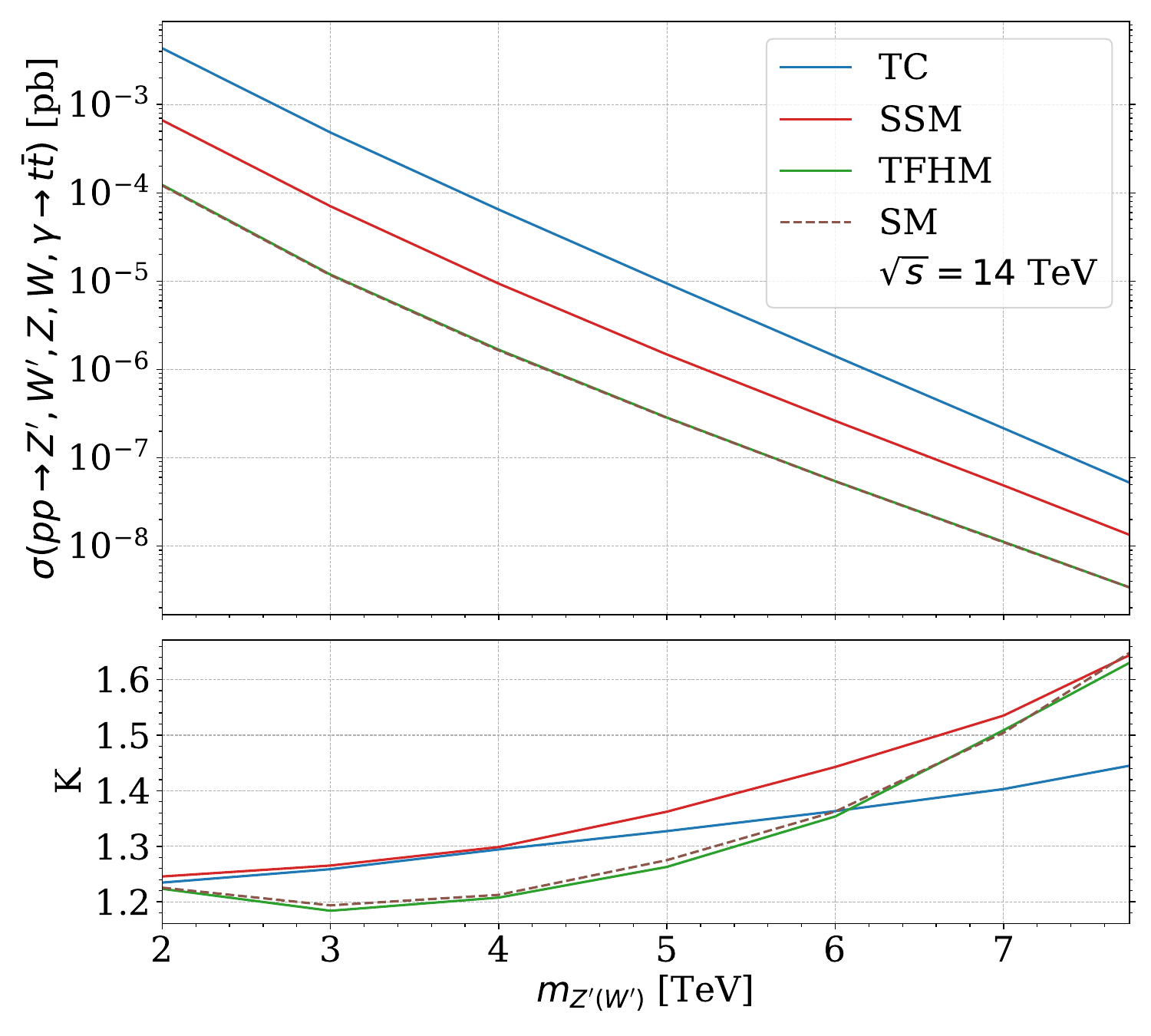}
   \includegraphics[width=0.45\textwidth]{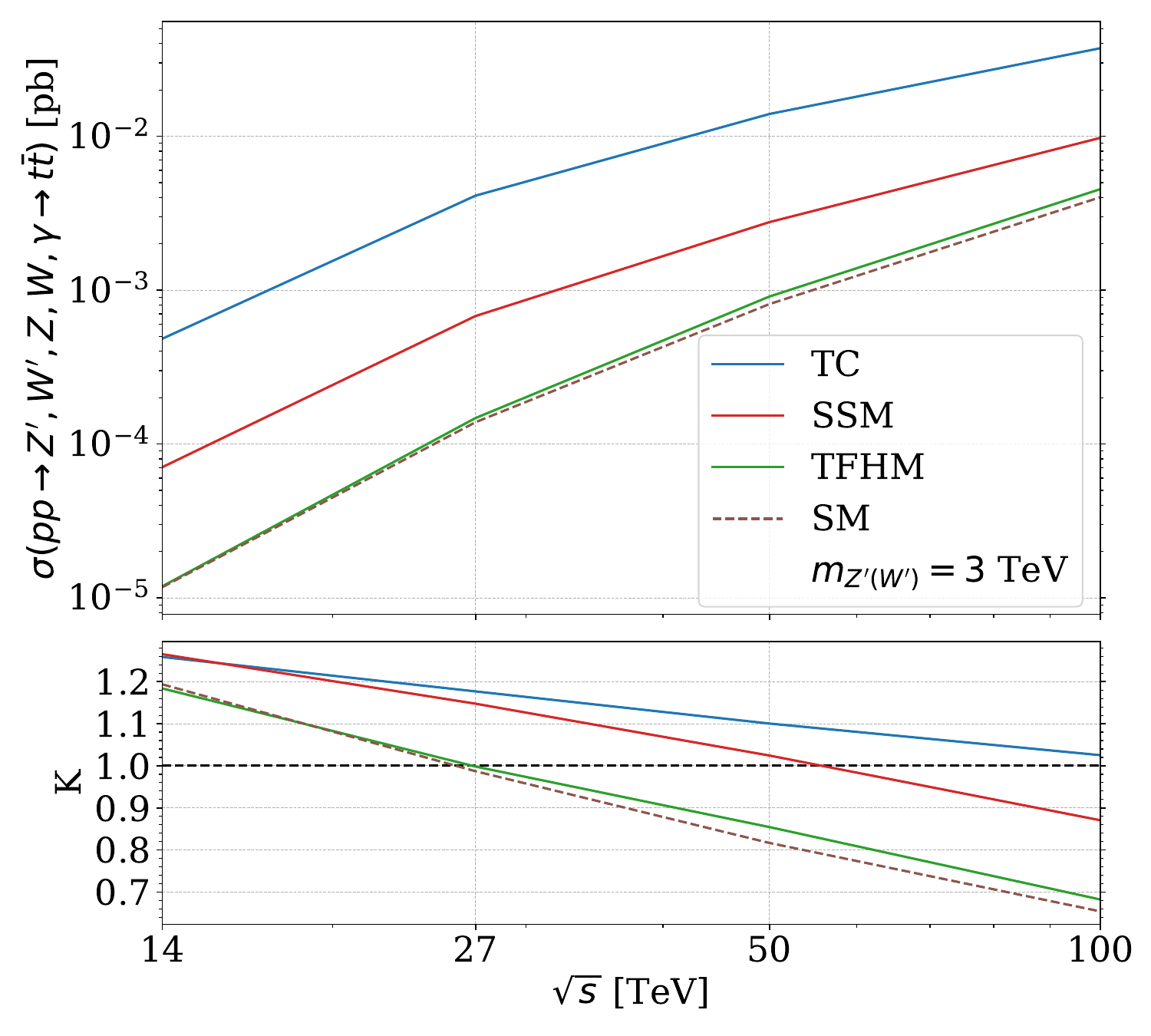}
	\caption{Fiducial cross section for EW $t\bar t$ production in SM, SSM, TC
	and THFM with an invariant mass cut $m_{t\bar t} \ge 0.75 m_{Z'}$ at NLO+PS
	(upper panels) and in ratio to LOPS (lower panels). Left panel: Cross section
	at $\sqrt{s}=14$ TeV as a function of $m_{Z'}$. Right panel: Cross section at
	$m_{Z'}=3$ TeV as a function of $\sqrt{s}$.} 
 \label{fig:01}
\end{figure}

In Fig.~\ref{fig:01}
we show the fiducial cross sections as a function of $m_{Z'}$ for $\sqrt{s}=14$
TeV and as a function of $\sqrt{s}$ for $m_{Z'} = 3$ TeV, on the left and right
respectively, in the SSM, TC and THFM models as well as in the SM.
The upper panels show the NLO+PS cross sections while the lower panels the
corresponding $K$-factors, i.e.~the ratio of NLO+PS over LO+PS cross sections.
First we note that the SM cross sections depend on the $m_{Z'}$ indirectly
through the invariant mass cut.
The TC cross sections are almost two orders of magnitude above the SM ones over
the entire $m_{Z'}$ and $\sqrt{s}$ ranges.
Similarly, the SSM cross sections are about one order magnitude larger than in
the SM.
The THFM curves overlap with the SM curve almost everywhere except for the
$\sqrt{s} \ge 50$ TeV where a slight excess over SM can be observed.
From this we conclude that the invariant mass cut for suppressing the SM
contribution to the EW top-pair production (background) designed to enhance the
$Z'$ mediated top-pair production (signal) is adequate in the cases of the
SSM and TC model.
However the THFM cross sections are very small even in the presence of this cut
and further cuts may be required for such production to be observable at the LHC.
The higher order QCD corrections are moderate and grow with the mass of
$m_{Z'}$.
With $\sqrt{s}$ the $K$-factor drops instead with higher order corrections for
SSM turning negative at $\sqrt{s} = 100$ TeV.

Now we turn our attention to the impact of newly included contributions on
the fiducial cross section.
In the SM and at LO when all the cuts are relaxed, the EW $t \bar{t}$ production
is dominated by the contribution due to $t$-channel $W$-boson exchange, which
is in fact about factor 6 larger than the resonant production.
Once the invariant mass cut is applied, both the resonant and non-resonant SM
contributions are reduced by about the same factor (by roughly 5 orders of
magnitude) and both drop well below the BSM signal.
However, the SM contributions respond quite differently to the fiducial cuts,
where the resonant production is reduced at the same rate as the BSM signal
(about 30\%), while the non-resonant production is reduced by over 90\%.

\begin{figure}
 \centering
   \includegraphics[width=0.45\textwidth]{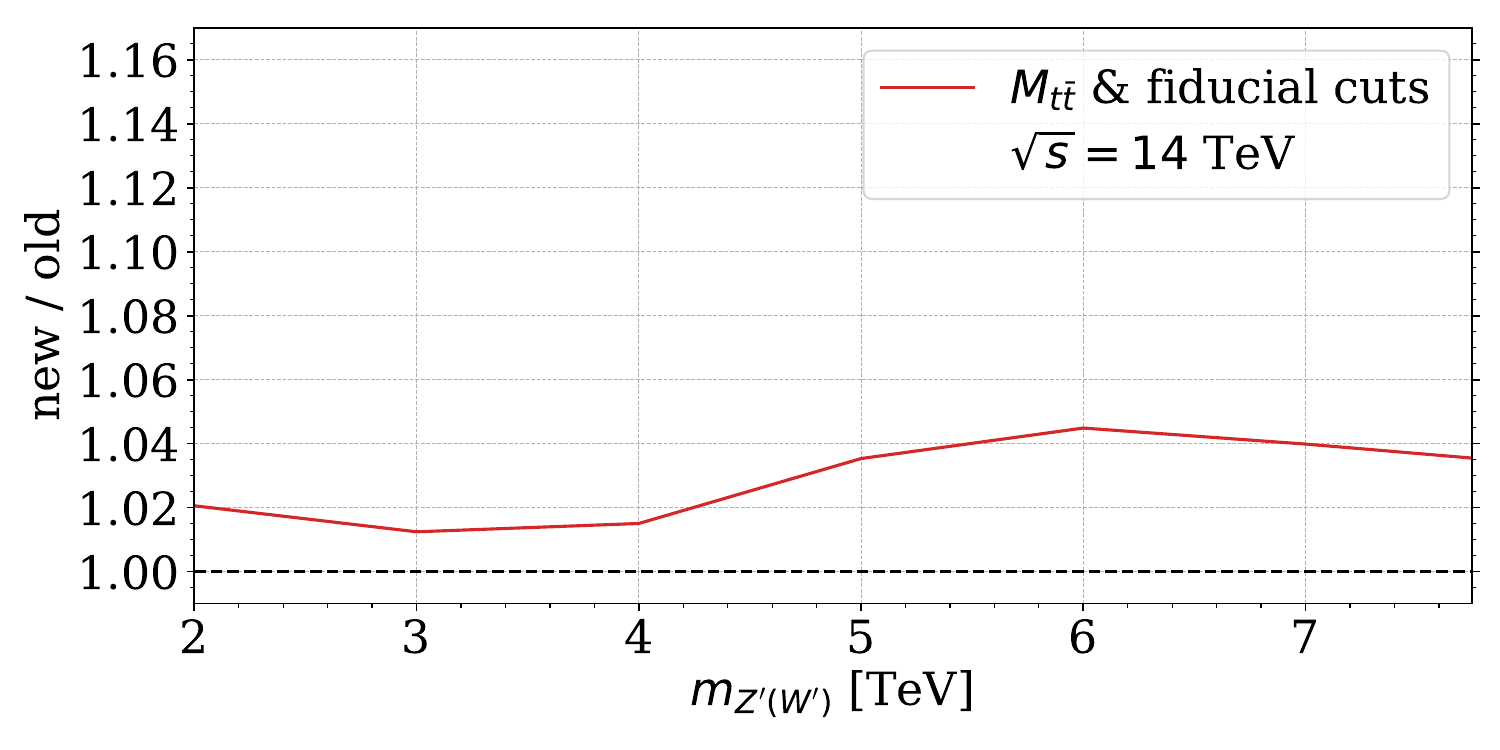}
   \includegraphics[width=0.45\textwidth]{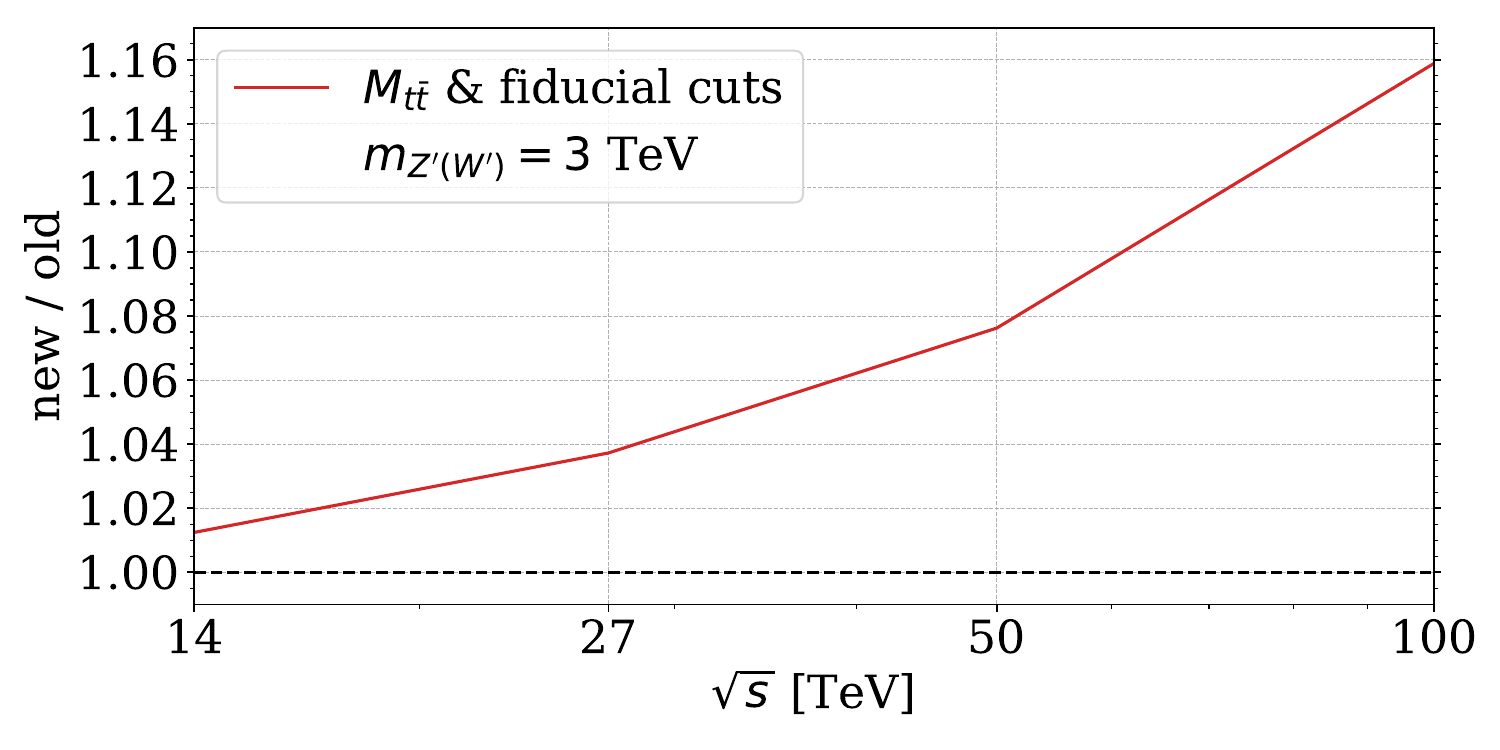}
	\caption{The ratio of fiducial cross section for EW $t\bar t$ production in
	SSM, with over without contributions due to $t$-channel $W$ boson exchange
	as a function of $m_{Z'}$. We set for $\sqrt{s}=14$ TeV and apply the usual invariant
	mass cut $m_{t\bar t} \ge 0.75 m_{Z'}$.}
 \label{fig:02}
\end{figure}

In Fig.~\ref{fig:02}
we show the ratio of the fiducial cross sections with and
without the new contributions.
We find that at $\sqrt{s} = 14$ TeV invariant mass and fiducial cuts turn the 
new contributions to a correction of a mere few percent with respect to our previous
predictions.
However, it can become quite important as we increase the centre-of-mass energy.

\section{Summary and conclusions \label{sec:conclusions}}

We presented a re-calculation of NLO QCD corrections to EW hadroproduction of
$t\bar{t}$ pairs in SM extensions with new heavy hypothetical neutral and
charged gauge bosons, $Z'$ and $W'$ bosons respectively.
In comparison to our previous calculation in Ref.~\cite{Bonciani:2015hgv},
we now allow flavour--non-diagonal $Z'$ couplings and include contributions
due to $t$-channel exchange of $W$ and $W'$ bosons. 
This significantly extends the range of SM extensions that can be considered.

We demonstrate this calculation on the SSM, TC and TFHM models for which we 
calculate NLOPS fiducial cross sections as a function of $Z'$ mass and
the centre-of-mass energy.
We also show the impact of the new non-resonant contributions which are mild at
$\sqrt{s} = 14$ TeV but can be relatively important at higher energies.
A more detailed and complete study is in preparation \cite{Altakach:2020ugg}.


\bibliographystyle{JHEP}
\bibliography{proceedings}

\end{document}